# A CONTINUOUS INJECTION PLASMA MODEL FOR THE X-RAY/RADIO KNOTS IN KPC-SCALE JETS OF AGN

S. Sahayanathan[1], R. Misra[2], A. K. Kembhavi[2] and C. L. Kaul[1]

## ABSTRACT

We consider the evolution of a spherically expanding plasma cloud, where there is continuous injection of non-thermal electrons. We compute the time dependent electron distribution and resultant photon spectra taking into account synchrotron, adiabatic and inverse Compton cooling. This model is different from previous works where, instead of a continuous injection of particles, a short injection period was assumed.

We apply this model to the radio/optical knots in the large scale jets of AGN, detected in X-rays by *Chandra* and find that the overall broadband spectral features can be reproduced. It is shown that for some sources, constraints on the X-ray spectral index (by a longer *Chandra* observation) will be able to differentiate between the different models. This in turn will put a strong constraint on the acceleration mechanism active in these sources.

*Subject headings:* Galaxies: active - galaxies: jets - X-rays: galaxies

## 1. Introduction

The detection of Kilo-parsec (kpc) scale jets with knots in several active galactic nuclei (AGN) by the *Chandra* observatory (Chartas *et al.* 2000; Tavecchio *et al.* 2000; Sambruna *et al.* 2002) has opened a new window on the nature of these phenomena. The X-ray emission from jets could be due to Comptonization of the optical/radio synchrotron photons i.e., synchrotron self-Compton (SSC), Comptonization of cosmic micro-wave background photons (IC/CMB), or an extension of the synchrotron radio/optical emission. Earlier observations by the *Einstein* and ROSAT satellites, having comparatively limited resolution and sensitivity, had not been generally able to differentiate amongst these mechanism (e.g. (Biretta *et al.* 1991; Harris & Stern 1987)). Based on the recent *Chandra* observation, it has been argued that the SSC interpretation would require large jet powers and magnetic fields much lower than the equipartition values (Tavecchio *et al.* 2000; Schwartz *et al.* 2000). A plausible attractive alternative is to interpret the X-ray emission as being due to IC/CMB or direct synchrotron, in which case a significantly smaller jet power and near-equipartition magnetic fields are required (Celotti *et al.* 2001; Pesce *et al.* 2001; Sambruna *et al.* 2002). For knots where the X-ray flux is greater than the extrapolation of the radio/optical spectra to X-ray wavelengths, the emission cannot be due to synchrotron emission and hence, the IC/CMB model is favored. On the other hand, for those knots whose X-ray flux is lower than this extrapolation, it is also possible to interpret the emission as being due to synchrotron from a non-thermal electron distribution with a high energy cut off. In fact, it has been argued (Pesce *et al.* 2001) that for the knots of 3C271, this is a more plausible explanation, since the IC/CMB model requires for this source an exceptionally large Doppler factor. In low-power FRI jets also, synchrotron origin of X-rays is currently accepted (Worrall *et al.* 2001). However, for the gigahertz-peaked spectrum (GPS) radio source PKS 1127-145, which is suspected to be a young FRI radio galaxy, the X-ray emission is probably due to IC/CMB (Siemiginowska *et al.* 2002).

[1]Nuclear Research Laboratory, Bhabha Atomic Research Center, Mumbai, India; sunder@apsara.barc.ernet.in

[2]Inter-University Center for Astronomy and Astrophysics, Post Bag 4, Ganeshkhind, Pune-411007, India; rmisra@iucaa.ernet.in

The interpretation of X-ray emission for some sources as being due to synchrotron emission is interesting, since in this case the X-rays are produced by particles having much higher energies (by a factor of $\approx 10^4$) than those emitting in the radio. The presence of high energy electrons puts constraints on the age of the knot: since these electrons cool efficiently, they are depleted in time unless they are replenished. The depletion leads to a time-dependent high energy cutoff in the non-thermal electron distribution which reflects as a high frequency exponential cutoff in the observed spectra. Since the observed -ray flux is less than the extrapolation of radio/optical flux in these sources, this exponential high frequency cutoff is constrained to be in the soft X-ray regime. This can be translated to a high energy cutoff in the electron distribution which in turn gives an estimate of the age of the knot. In this model, a knot is formed when a short duration acceleration process enhances the non-thermal electron density in a jet. These non-thermal electrons move with a bulk speed $v \approx c$ along the jet. Thus, from the age of the knot, one can determine the location in the jet of the short acceleration process. The distance from the central object and the short duration ( $<<$ the age of the knot) naturally puts strong constraints on any models of the acceleration process.

The model outlined above, may be confirmed (or ruled out) by future long duration *Chandra* observations. The model predicts that the X-ray spectra of the knots should be exponential (i.e. steep). The photon spectral slope measured during preliminary short duration observations of 3C 371 (Pesce *et al.* 2001), is around $\Gamma = 1.7 \pm 0.4$, in apparent contradiction to this prediction. However, longer observations are required to confirm this result. Moreover, the model requires the coincidence that the age of the knot be equal to the time required for X-ray emitting electrons to cool. A larger survey of X-ray jets have to be sampled to confirm whether this is statistically plausible. In this model, it is assumed that the acceleration time-scale is short (i.e. $t_{acc} << 10^{10}$ sec) and the possible expansion of the plasma is not taken into account. The acceleration mechanism of these knots is largely unknown, but if the acceleration is due to internal shocks (e.g. (Spada *et al.* 2001)), then the acceleration time scale $t_{acc} \approx R_s/c$, where $R_s$ is a typical shell size. If $R_s \approx$ kpc, then $t_{acc}$ could be as large as $\approx 10^{11}$ secs.

In this work, we present an alternative model where the knot is assumed to be a uniform expanding sphere with continuous injection of non-thermal particles. The time-dependent electron distribution and the resultant spectra are computed taking into account synchrotron, IC/CMB and adiabatic cooling due to the expansion of the sphere. Since there is continuous injection of particles, there is no high energy cut-off but, instead, there is a time dependent break in the electron distribution where synchrotron/adiabatic cooling is important. Our motivation here is to show that this model also explains the observed spectra and can be distinguished from one time injection models, by future observational constrains on the X-ray spectral index.

In the next section, the model is described and the predicted spectral energy distributions are compared with observations. §3 summarizes and discusses the main results. Throughout this work, $H_o = 75$ km s$^{-1}$ Mpc$^{-1}$ and $q_0 = 0.5$ are adopted.

## 2. Spherically expanding plasma model

We consider a plasma cloud moving relativistically along the jet with a bulk Lorentz factor $\Gamma$. In the rest frame, it is assumed that the plasma uniformly occupies an expanding sphere of radius $R(t) = R_o + \beta_{exp}ct$, where $R_o$ is the initial size of the sphere and $\beta_{exp} << 1$. At $t = 0$, there are no non-thermal particles in the system. A continuous and constant particle injection rate [ i.e. number of particles injected per unit time] for $t > 0$ is assumed, with a power-law distribution of energy,

$$Q(\gamma)d\gamma = K\gamma^{-p}d\gamma \quad \text{for} \quad \gamma > \gamma_{min} \qquad (1)$$

where $\gamma$ is the Lorentz factor of the electrons. The kinetic equation describing the evolution of the total number of non-thermal particles in the system, $N(\gamma, t)$, is

$$\frac{\partial N(\gamma, t)}{\partial t} + \frac{\partial}{\partial \gamma}[P(\gamma, t)N(\gamma, t)] = Q(\gamma) \tag{2}$$

$P(\gamma, t)$ is the cooling rate given by

$$P(\gamma, t) = -(\dot{\gamma}_S(t) + \dot{\gamma}_{IC}(t) + \dot{\gamma}_A(t)) \tag{3}$$

where $\dot{\gamma}_S(t), \dot{\gamma}_{IC}(t)$ and $\dot{\gamma}_A(t)$, the cooling rates due to synchrotron, inverse Compton of the CMB and adiabatic cooling, respectively, are given by

$$\dot{\gamma}_S(t) = \frac{4}{3}\frac{\sigma_T}{m_e c}\frac{B^2(t)}{8\pi}\gamma^2 \tag{4}$$

$$\dot{\gamma}_{IC}(t) = \frac{16}{3}\frac{\sigma_T}{m_e c^2}\Gamma^2 \sigma T_{cmb}^4(z)\gamma^2 \tag{5}$$

$$\dot{\gamma}_A(t) = \frac{\beta_{exp} c \gamma}{R(t)} \tag{6}$$

Here, the evolving magnetic field is parameterized to be $B(t) = B_0(R(t)/R_0)^m$ and $T_{cmb}(z) = 2.73(1+z)$ is the temperature of the CMB at the redshift $z$ of the source. Note that the time $t$ and other quantities in the above equations are in the rest frame of the plasma.

In this work, Eqn (2) has been solved numerically for $N(\gamma, t)$ using the technique given by (Chang & Cooper 1970) and the resultant synchrotron and inverse Compton spectra are computed at an observing time $t = t_o$. As shown by (Dermer 1995), since the CMB radiation is not isotropic in the rest frame of the plasma, the inverse Compton spectrum is also beamed. Finally the flux at the earth is computed taking into account the Doppler boosting (e.g. (Begelman & Blandford 1984)), characterized by the Doppler factor $\delta \equiv [\Gamma(1 - \beta \cos\theta)]^{-1}$, where $\beta c$ is the bulk velocity and $\theta$ is the angle between the jet and the line of sight.

While the total non-thermal particle distribution has to be computed numerically, a qualitative description is possible by comparing cooling timescales with the observation time $t_o$. The cooling time-scale due to synchrotron and inverse Compton cooling at a given time $t$ and Lorentz factor $\gamma$, is $t_c(t, \gamma) \approx \gamma/(\dot{\gamma}_S + \dot{\gamma}_{IC})$. Then, $\gamma_c$, defined as the $\gamma$ for which this cooling time scale is equal to the observation time, $t_c(t_o, \gamma_c) \approx t_o$, becomes,

$$\gamma_c \approx \frac{m_e c}{\sigma_T}\frac{t_o^{-1}}{[\frac{B^2(t_o)}{8\pi} + \Gamma^2 4 c \sigma T_{cmb}^4(z)]} \tag{7}$$

The adiabatic cooling time-scale also turns out to be $\approx t_o$, since $t_a \approx R(t)/\beta_{exp}c \approx t_o$ for $R_o \ll R(t_o)$. Thus, the non-thermal particle distribution at time $t = t_o$ can be divided into three distinct regions:

(i) $\gamma \ll \gamma_c$: In this regime, radiative cooling is not important and $N(\gamma, t_o) \approx K\gamma^{-p}t_o$. The corresponding spectral index for both synchrotron and inverse Compton is $\alpha = (p-1)/2$.

(ii) $\gamma \gg \gamma_c$: In this regime, either synchrotron or inverse Compton cooling is dominant and $N(\gamma, t_o) \propto \gamma^{-(p+1)}$. The corresponding spectral index for both synchrotron and inverse Compton is $\alpha = p/2$.

(iii) $\gamma \approx \gamma_c$ In this regime, either synchrotron or inverse Compton cooling as well as adiabatic cooling are important, and the spectral slope is in the range $(p-1)/2 > \alpha > p/2$.

The computed spectra depend on the following ten parameters: the observation time $t_o$, the magnetic field at the time of observation $B_f = B(t = t_o)$, the magnetic field variation index $m$, the radius of the knot at the time of observation, $R_f = R(t = t_o)$, the index $p$, the minimum Lorentz factor $\gamma_{min}$, the Doppler factor $\delta$, the bulk Lorentz factor $\Gamma$, expansion velocity $\beta_{exp}$ and the normalization of the injection rate $K$. On the other hand, there are only three observational points namely, the radio/optical and X-ray fluxes. Clearly the parameters are under-constrained and it is not possible to extract meaningful quantitative estimates. However, the motivation here is to show that this model can explain the observed data with reasonable values of the above parameters.

In Figure 1, the computed spectra are compared with the data for different knots for four sources. The values of the parameters used are tabulated in Table 1. The injected power in non-thermal particles in the rest frame is

$$P_{inj} = \int_{\gamma_{min}}^{\infty} (\gamma m_e c^2) Q(\gamma) d\gamma = K \frac{m_e c^2}{p-2} \gamma_{min}^{-(p-2)} \tag{8}$$

while the power in the jet can be approximated to be (e.g. (Celotti *et al.* 1997)),

$$P_{jet} = \pi R^2 \Gamma^2 \beta c (U_p + U_e + U_B) \tag{9}$$

where $U_p$, $U_e$ and $U_B$ are the energy densities of the protons, electrons and the magnetic field, respectively. Here, it has been assumed that the protons are cold and the number of protons is equal to the number of electrons. The jet power ranges from $10^{46}$ to $2 \times 10^{48}$ ergs sec$^{-1}$, while the injected power is generally three orders of magnitude lower. This means that the non-thermal acceleration process is inefficient and most of the jet power is expected to be carried to the lobes. The magnetic field $B_f$ is nearly equal to the equipartition values.

Like the results obtained by (Sambruna *et al.* 2002) and (Pesce *et al.* 2001), the X-ray emission from the knots in 3C371 and Knot A of 1136-135 are identified as being due to synchrotron emission. However, in this case the predicted X-ray spectral index is $\alpha_X = \alpha_R + 1/2$ instead of being exponential. Note that this relation between the spectral indicies is independent of the parameters used to fit the data. For the rest of the sources, like the earlier results (Sambruna *et al.* 2002), the X-ray emission is attributed to IC/CMB. However, for some of the sources the optical spectral index is now $\alpha_O = \alpha_R + 1/2$, instead of being exponential.

## 3. Summary and Discussion

In this work, it is shown that the observed radio, optical and X-ray fluxes of knots in kpc scale jets in AGN, can be explained within a framework of a model where there is continuous injection of non-thermal particles into an expanding spherical plasma. This interpretation can be confirmed (or ruled out) vis-a-vis one-time injection models, by future measurements of the radio $\alpha_R$, optical $\alpha_O$ and X-ray $\alpha_X$ spectral indicies. In particular, the following cases are possible:

(i) $\alpha_R \approx \alpha_X$: In this case, the X-ray emission is probably due to IC/CMB. Both the continuous injection and one-time injection models are equally viable.

(ii) $\alpha_X \approx \alpha_R + 1/2$ : In this case, the X-ray emission would be due to synchrotron emission from electrons in the cooling dominated region. The continuous injection scenario will be favored.

(iii) $\alpha_X > \alpha_R + 1/2$ : In this case, when the X-ray emission is exponentially decreasing, it should be attributed to the high energy cutoff in the electron distribution. The one-time injection scenario will be favored.

(iv) $\alpha_X < \alpha_R$ : In this case, when the X-ray emission is exponentially increasing, it should be attributed to the low energy cutoff ($\gamma_{min}$) in the electron distribution and the X-ray emission should be due to IC/CMB. Both the continuous injection and one-time injection models are equally viable.

Similar arguments can be put forth for the optical spectral index $\alpha_O$ as compared to the radio. It should be noted that for some older systems the one time-injection would be the natural scenario, while for younger systems the continuous one would be more probable. The technique described above will be able to differentiate between the two and a generic constraint on the acceleration time-scales and typical age of the knots may be obtained. A generic model where the injection rate decays in time, may then be used to fit the observations. The measurement of spectral indicies at different wave-lengths will also reduce the number of unconstrained parameters in the model fitting, leading to reliable estimates of the system parameters.

Constraints on the acceleration time-scale and the age of the knots would be the first step toward understanding the driving mechanism active in these sources. It will then be possible to compare these time-scales with theoretical results from analytical or numerical computations, which may finally lead to an understanding of the origin and nature of jets in AGN.

Table 1. Parameters for model fitting

| Source/Knot | $B_f$ | $\gamma_{min}$ | $p$ | $t_o$ | $\delta$ | $\Gamma$ | $\beta_{exp}$ | $P_{inj}$ | $P_{jet}$ | $B_f/B_{equ}$ |
|---|---|---|---|---|---|---|---|---|---|---|
| 1136-135 A | 0.9 | 2.0 | 2.4 | 0.2 | 5 | 5 | 0.8 | 44.7 | 47.5 | 0.4 |
| B | 4.0 | 20.0 | 2.9 | 9 | 5 | 5 | 0.1 | 44.2 | 47.9 | 0.5 |
| 1150+497 A | 2.5 | 30.0 | 2.85 | 9 | 5 | 3.5 | 0.1 | 44.2 | 47.3 | 0.4 |
| B | 4.3 | 30.0 | 3.3 | 9 | 5 | 3.5 | 0.1 | 44.1 | 47.3 | 0.75 |
| 1354+195 A | 1.7 | 40.0 | 3.0 | 9 | 3.5 | 2 | 0.1 | 45.8 | 48.4 | 0.04 |
| B | 8.0 | 25.0 | 3.2 | 9 | 3.5 | 2 | 0.1 | 44.7 | 47.5 | 0.63 |
| 3C 371 A | 1.3 | 10.0 | 2.4 | 12 | 3.5 | 3.5 | 0.1 | 42.8 | 46.4 | 0.8 |
| B | 1.0 | 10.0 | 2.4 | 1 | 3.5 | 3.5 | 0.5 | 43.5 | 46.0 | 0.9 |

Note. — Columns:- 1: source and knot name taken from (Pesce *et al.* 2001) for 3C371 and the rest from (Sambruna *et al.* 2002). 2: magnetic field in units of $10^{-5}$ Gauss at the observation time, $B_f = B(t = t_o)$. 3: minimum Lorentz factor $\gamma_{min}$. 4: non-thermal injection index $p$. 5: observation time $t_o$ in units of $10^{11}$ sec. 6;: Doppler factor $\delta$. 7: bulk Lorentz factor $\Gamma$. 8: log of the injected power in ergs s$^{-1}$ $P_{inj}$. 9: log of the total jet power in ergs s$^{-1}$ $P_{jet}$. 10: ratio of the magnetic field to the equipartition value. For all cases, the magnetic field variation index $m$ and the size of the source at $t = t_o$ is fixed at 1.5 and $5 \times 10^{21}$ cm, respectively.

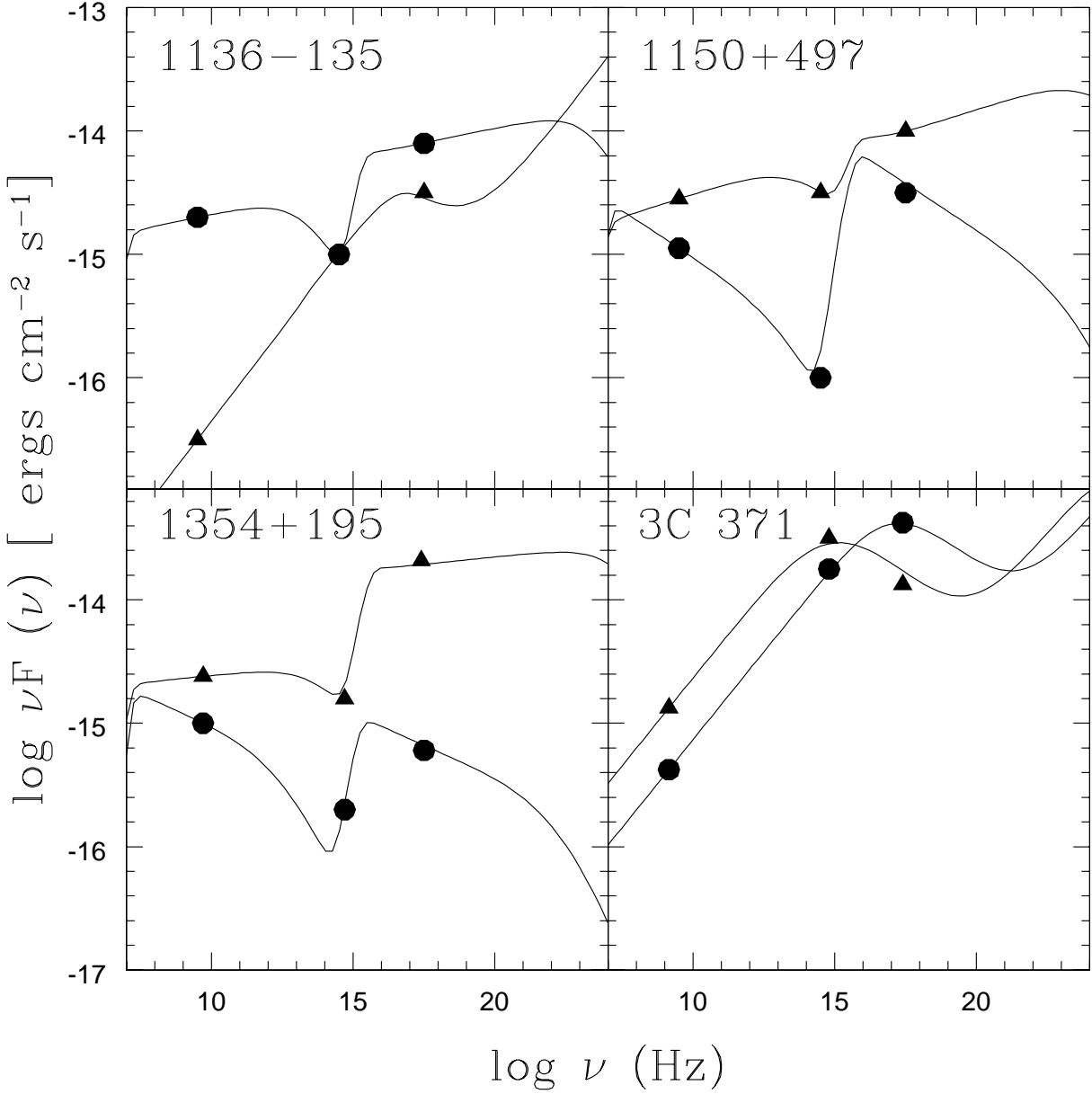

Fig. 1.— The observed fluxes in radio, optical and X-ray compared with model spectra using parameters given in Table 1. The data for 3C 371 are taken from (Pesce *et al.* 2001), while the rest are taken from (Sambruna *et al.* 2002). Triangles (dots) correspond to knot A (B). Errors are typically 30% or larger.